\pdfoutput=1

\documentclass[conference,compsoc]{IEEEtran}
\IEEEoverridecommandlockouts

\usepackage[normalem]{ulem} 

\usepackage{enumerate} 

\usepackage{color}

\ifCLASSOPTIONcompsoc
  \usepackage[nocompress]{cite}
\else
  \usepackage{cite}
\fi
\ifCLASSINFOpdf
  \usepackage[pdftex]{graphicx}
\else
\fi
\ifCLASSOPTIONcompsoc
 \usepackage[caption=false,font=footnotesize,labelfont=sf,textfont=sf]{subfig}
\else
 \usepackage[caption=false,font=footnotesize]{subfig}
\fi
\begin{document}
%
\title{Evaluating Dynamic File Striping For Lustre}

\author{\IEEEauthorblockN{Joel Reed \& Jeremy Archuleta}
\IEEEauthorblockA{Computational Science \& Engineering Division\\
Oak Ridge National Laboratory\\
Oak Ridge, Tennessee, U.S.A\\
\{reedjw, archuletajs\}@ornl.gov}
\and
\IEEEauthorblockN{Michael J. Brim \& Joshua Lothian}
\IEEEauthorblockA{Computer Science \& Mathematics Division\\
Oak Ridge National Laboratory\\
Oak Ridge, Tennessee, U.S.A\\
\{brimmj, lothian\}@ornl.gov}
\thanks{
This research was funded by the DoD-HPC program at Oak Ridge National Laboratory. This manuscript has been authored by UT-Battelle, LLC under Contract No. DE-AC05-00OR22725 with the U.S. Department of Energy. The United States Government retains and the publisher, by accepting the article for publication, acknowledges that the United States Government retains a non-exclusive, paid-up, irrevocable, world-wide license to publish or reproduce the published form of this manuscript, or allow others to do so, for United States Government purposes. The Department of Energy will provide public access to these results of federally sponsored research in accordance with the DOE Public Access Plan (http://energy.gov/downloads/doe-public-access-plan).
}}

\IEEEtitleabstractindextext{%
\begin{abstract}
We define dynamic striping as the ability to assign different Lustre striping characteristics to contiguous segments of a file as it grows. In this paper, we evaluate the effects of dynamic striping using a watermark-based strategy where the stripe count or width is increased once a file's size exceeds one of the chosen watermarks. To measure the performance of this strategy we used a modified version of the IOR benchmark, a netflow analysis workload, and the \texttt{blastn} algorithm from NCBI BLAST.  The results indicate that dynamic striping is beneficial to tasks with unpredictable data file size and large sequential reads, but are less conclusive for workloads with significant random read phases. 
\end{abstract}

\begin{IEEEkeywords}
Lustre, file striping, data analytics.
\end{IEEEkeywords}}

\maketitle

\IEEEdisplaynontitleabstractindextext

%
\IEEEpeerreviewmaketitle

\section{Introduction}

Historically, Lustre development has focused on improving the parallel I/O performance of scientific workloads. For these workloads, reading input datasets and writing result datasets (or intermediate checkpoint datasets) using coordinated parallel reads and writes are the predominant I/O patterns. Typically, data is accessed sequentially and the size of the dataset and expected number of processes performing I/O is known a priori. Using this knowledge, scientific applications can choose values for the Lustre file storage parameters, such as stripe count and stripe width, that will yield good performance by balancing load across servers and effectively utilizing storage network bandwidth. 

Recently, data analytic workloads have become more prevalent, and these workloads have I/O patterns and storage requirements that are quite different from traditional scientific applications. Data analytics applications often use a sequential input scan to generate an index of an extremely large dataset, and then perform various modeling, clustering, or statistical analysis tasks that use the index to correlate subsets of the data. The I/O pattern generated by these analysis tasks is read-intensive and mostly random from the perspective of the storage system. The number of processes used by an analytic workload can be independent of the dataset size, and may differ from one run to the next due to resource availability. Further complexity is introduced when the input to an analytic workload comes from a streaming data source, such as packet flows on a network, in which case it is impossible to predict the size of data to be stored for a given collection interval.

In general, Lustre's requirement for statically specifying the striping parameters for a file at the time of its creation is problematic when any one or more of the following are not known at creation time: 1) the final file data size, 2) the future read access patterns, or 3) the number of processes that will read the file. Currently, Lustre users can attempt to workaround this issue by assigning a wide stripe count in anticipation of a very large file. When the file is indeed very large, data will be spread across many servers, and those servers can potentially service many clients with good utilization of storage bandwidth and the network. If, however, the file is only accessed by a few clients, then read performance will be limited as each client will need to communicate with several servers. When the file turns out not to be very large, a wide-striping simply creates unnecessary contention at servers and results in poor utilization. Our goal is to avoid such bad striping parameter selections by eliminating the requirement for specifying a static striping at file creation.

We define the phrase ``dynamic striping'' as the ability to apply different striping parameters to contiguous segments of a file as it grows. A key requirement for supporting dynamic striping is \emph{Composite File Layouts} as proposed in~\cite{lfs_composite_layout}, which provides the necessary meta-data for per-segment striping. As a first evaluation of dynamic striping, we have chosen a simple watermark-based layout where a file starts with a relatively narrow stripe count that increases once the file size exceeds a predetermined watermark size. Several levels of watermark can be defined that further increase the stripe count. 
Our hypothesis is that data stored in this fashion should increase the read performance of analytic workloads where the file size is not known in advance.

\section{Experimental Method}

\subsection{Synthetic Dynamic Striping}

Given an analysis task where the final data size is initially unknown, read performance may be improved by adding dynamic striping support to the Lustre file system. 
To evaluate dynamic striping without modifying the Lustre client or server software, we simulate its effects by splitting files into segments at predetermined watermarks. 
Segments are stored as files in directories with different striping configurations. 
For example, if there are two watermarks at 1~GB and 10~GB and dynamic striping is to be applied to a 14~GB file, then three file sections are created from the original file representing:

\begin{itemize}
  \item the first 1~GB of the file,
  \item the next 9~GB of file data between offsets 1~GB and 10~GB, and
  \item the remaining 4~GB of the file beyond the 10~GB offset
\end{itemize}

Each file is then written to a specific directory having a specific stripe count and stripe width, as exemplified in Table~\ref{tab:synthetic_striping}.
If the analysis code does not support multiple files, it is modified to use a simple I/O interposition layer that makes the collection of file segments appear as a single file.
Table~\ref{tab:directory_types} lists the directory types and stripe count-stripe width combinations that were used during our evaluation.

\begin{table}[t]
\centering
\caption{Example of synthetic striping with two watermarks}
\label{tab:synthetic_striping}
\begin{tabular}{llll}
File Name & Data Range & Stripe Count & Stripe Width \\
\hline
4ost-1mb/part-00 & 0-1~GB & 4 & 1~MB \\
8ost-2mb/part-01 & 1-10~GB & 8 & 2~MB \\
16ost-4mb/part-02 & 10-14~GB & 16 & 4~MB
\end{tabular}
\end{table}

\begin{table}[t]
\centering
\caption{Directory Types with Stripe Width-Stripe Count}
\label{tab:directory_types}
\begin{tabular}{l||llllllllll}
Dir. Type & A & B & C  & D  & E & F  & G  & H & I  & J  \\
\hline
Stripe Count   & 4 & 8 & 16 & 64 & 4 & 16 & 64 & 4 & 16 & 64 \\
Stripe Width   & 1 & 1 & 1  & 1  & 2 & 2  & 2  & 4 & 4  & 4  \\
\end{tabular}
\end{table}

\subsection{Performance Measures}

We evaluated our synthetic dynamic striping to explore its effects on Lustre using three performance measures: 1) a modified version of the IOR benchmark, 2) a netflow analysis workload, and 3) the \texttt{blastn} algorithm from NCBI BLAST.

\subsubsection{IOR Benchmark}

IOR is a parallel file system benchmark using MPI for process synchronization and allowing various interfaces and access patterns to be evaluated~\cite{ior}. The IOR benchmark was modified to take an ordered list of directory paths and associated segment sizes, and split the file it creates across the directories abiding by the provided segment size. 

\begin{table}[h]
\centering
\caption{IOR Benchmarks}
\label{tab:ior_benchmarks}
\begin{tabular}{ll}
Experiment Name & Striping Pattern \\
\hline
IOR.1 & Entire file in \texttt{A} \\
IOR.2 & Entire file in \texttt{B} \\
IOR.3 & Entire file in \texttt{C} \\
IOR.4 & 0-1~TB in \texttt{A}, remainder in \texttt{B} \\
IOR.5 & 0-1~TB in \texttt{A}, remainder in \texttt{C} \\
IOR.6 & 0-1~TB in \texttt{A}, 1-2~TB in \texttt{B}, remainder in \texttt{C} \\
\end{tabular}
\end{table}

The IOR benchmark was executed using 64 tasks spread evenly across 16 nodes, with a block size setting of 64 GB, resulting in a total file size of 4 TB. Between each run the page cache on each node was dropped and a large, unrelated file was read from the Lustre filesystem to try and minimize the cache effect. Utilizing the Lustre directory types described in Table~\ref{tab:directory_types}, the IOR benchmarks listed in Table~\ref{tab:ior_benchmarks} were each run five times.

\subsubsection{netflow}

RFC 2722 defines a netflow as an artificial logical equivalent to a call or connection~\cite{rfc2722}. Essentially, a netflow is a representation of a collection of network packets making up a single protocol connection between a pair of computers, usually limited to a time interval.
The netflow analysis workload takes a collection of netflow data and produces a behavior model for each internal IP address. This analysis consists of two I/O phases. In phase 1, each task gets a segment of the input data to read and index, and in phase 2, each task gets a list of record offsets to read and create a model from. Phase 1 can be characterized as a set of large sequential reads and phase 2 as a large number of small, random reads. 

\begin{figure}[t]
\centering
  \includegraphics[width=1.0\columnwidth]{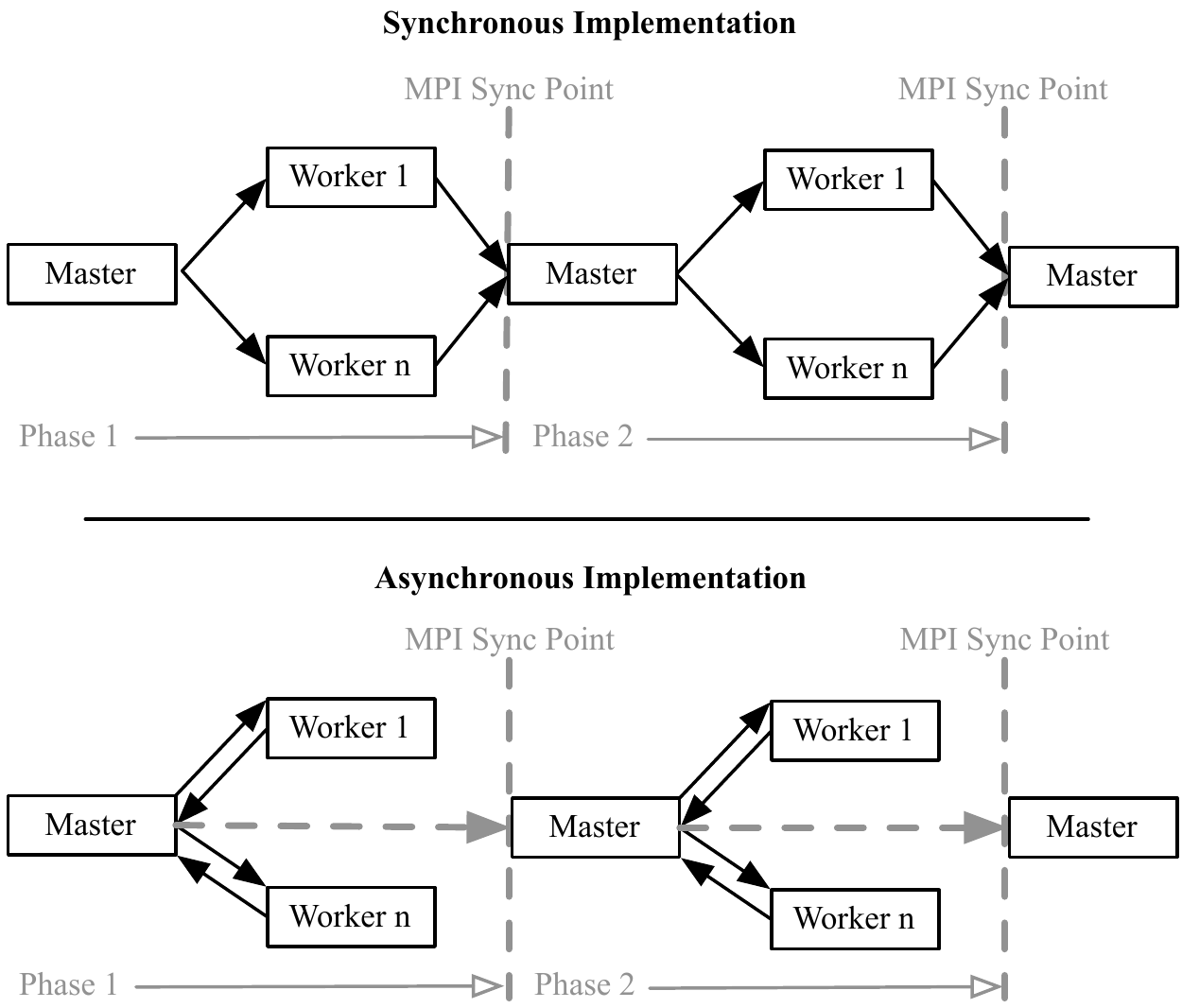}
\caption{Netflow workload \- synchronous vs. asynchronous design}
\label{fig:netflow}
\end{figure}

This workload was implemented in both a synchronized variation and an asynchronous variation. In the synchronous variation, each phase 1 task received an equal portion of the input data and each phase 2 task was given an approximately even sized set of the record offsets to process. In the asynchronous variation, each phase 1 task was given a fixed size segment to process and was given another when ready, and each phase 2 task was given a set of offsets representing a single model and was given another when ready (see Figure~\ref{fig:netflow}).

The netflow workload was modified to operate on several netflow data file segments, spread across several directories, as if they were a single file. The input data is 55 GB of netflows. The netflow workload was executed as 128 tasks spread evenly across 16 nodes. To minimize the cache effect, each node's page cache was dropped and a large, unrelated file was read from the Lustre filesystem.

Utilizing the Lustre directory types described in Table~\ref{tab:directory_types}, the 
netflow benchmarks listed in Table~\ref{tab:netflow_benchmarks} were each run three times using the synchronous variation and three times using the asynchronous variation. For each of the 2 phases, the span of time from the first I$/$O operation started to the time that the final I$/$O operation completed was collected. Also, for each phase, the composite throughput was collected.

\begin{table}[h]
\centering
\caption{netflow Benchmarks}
\label{tab:netflow_benchmarks}
\begin{tabular}{ll}
Experiment Name & Striping Pattern \\
\hline
netflow.1 & Entire file in \texttt{A} \\
netflow.2 & Entire file in \texttt{B} \\
netflow.3 & Entire file in \texttt{C} \\
netflow.4 & 0-10~GB in \texttt{A}, remainder in \texttt{B} \\
netflow.5 & 0-10~GB in \texttt{A}, remainder in \texttt{C} \\
netflow.6 & 0-10~GB in \texttt{A}, 10-20~GB in \texttt{B}, remainder in \texttt{C} \\
\end{tabular}
\end{table}

\subsubsection{BLAST}

The Basic Local Alignment Search Tool (BLAST)\cite{BLAST} and its constituent algorithms are developed by the National Center for Biotechnology Information (NCBI) at the National Institutes of Health (NIH). Similar to the {\tt grep} command, individual BLAST algorithms take a query file consisting of DNA or protein sequences and search for similar sequences within a target database of DNA or protein sequences. An unknown query sequence that closely matches a known target sequence helps a bioinformatician understand the functionality and lineage of the unknown query sequence.

\begin{figure*}[!t]
\centering
\subfloat[Phase I: Identify word hits (i.e., seeds)]{\includegraphics[width=2.25in]{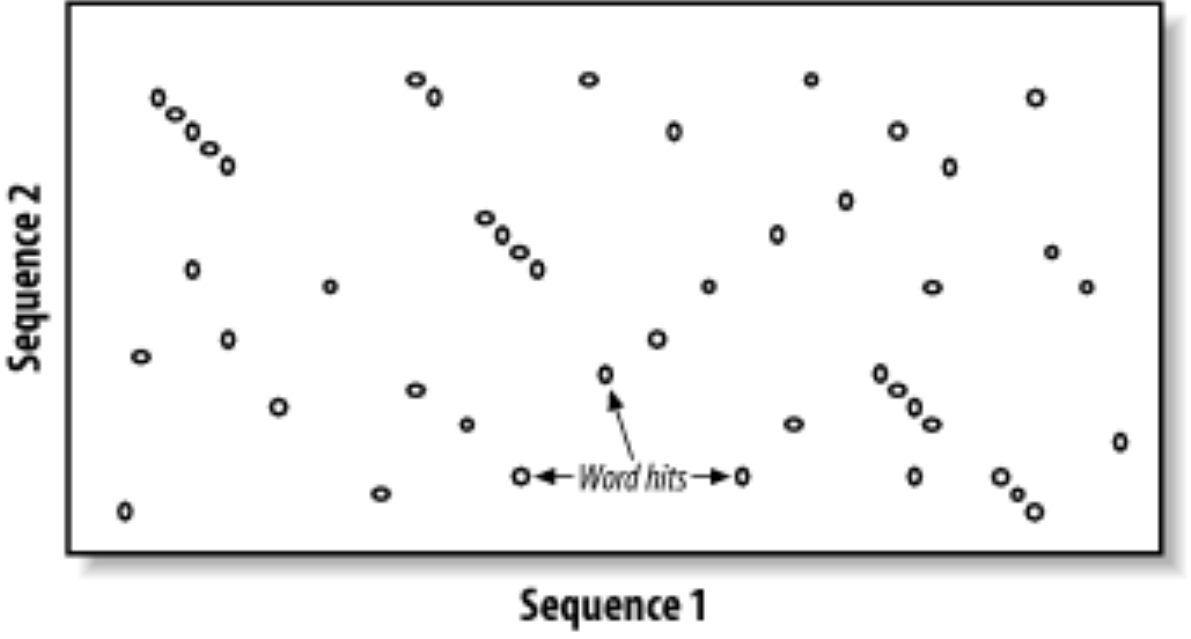}%
\label{fig:blast_seed}}
\hfil
\subfloat[Phase II: Combine ``close'' clusters]{\includegraphics[width=2.25in]{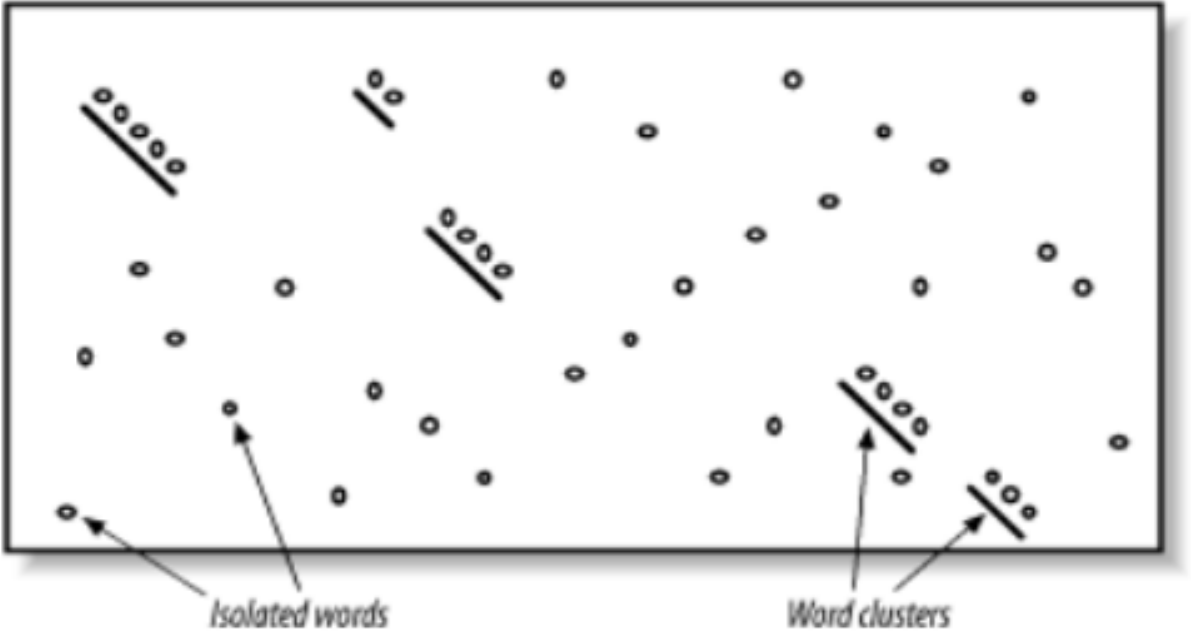}%
\label{fig:blast_combine}}
\hfil
\subfloat[Phase III: Extend each match]{\includegraphics[width=2.25in]{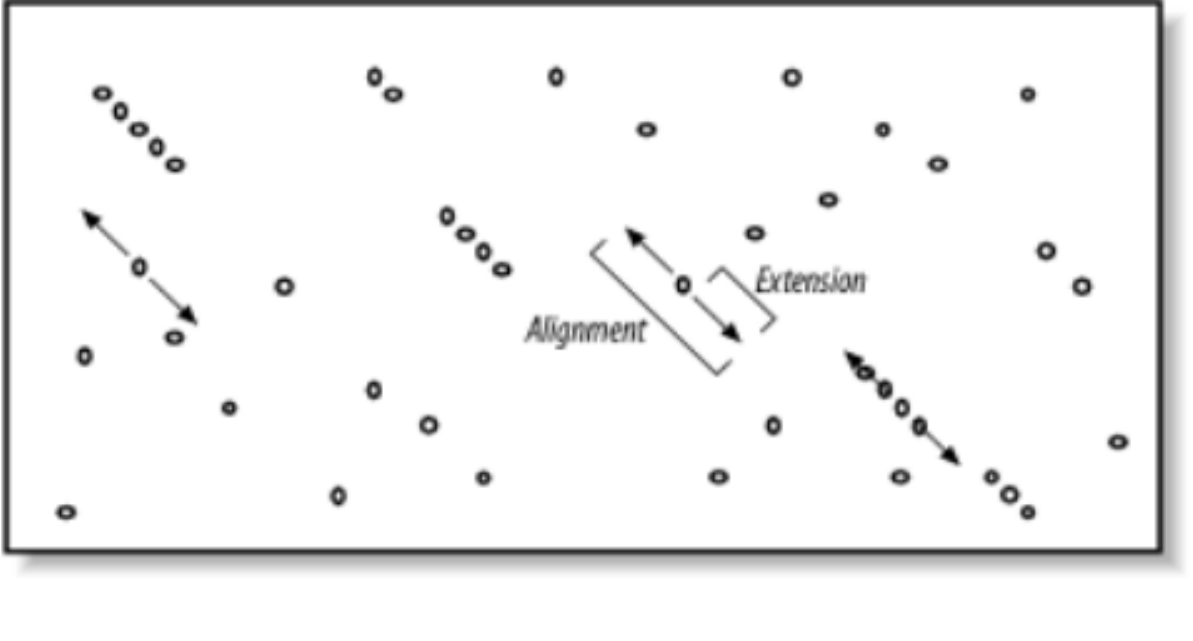}%
\label{fig:blast_extend}}
\caption{Three phases of the BLAST algorithm.}
\label{fig:blast_phases}
\end{figure*}

Due to the computational complexity of sequence matching, BLAST uses an heuristic to dramatically reduce the search space. The heuristic is implemented in three phases where the first phase identifies seeds where subsequences of the query and target sequences match with a score of at least $T$~(Figure \ref{fig:blast_seed}). The second phase combines matches that are within $G$ letters of each other~(Figure \ref{fig:blast_combine}). The final phase extends each match until the score for the alignment drops below $S$~(Figure \ref{fig:blast_extend}).

\begin{table}[t]
\centering
\caption{BLAST Benchmarks}
\label{tab:blast_benchmarks}
\begin{tabular}{ll}
Experiment Name & Striping Pattern \\
\hline
blast.1  & Entire file in \texttt{A} \\
blast.2  & Entire file in \texttt{C} \\
blast.3  & Entire file in \texttt{D} \\
blast.4  & Entire file in \texttt{E} \\
blast.5  & Entire file in \texttt{F} \\
blast.6  & Entire file in \texttt{G} \\
blast.7  & Entire file in \texttt{H} \\
blast.8  & Entire file in \texttt{I} \\
blast.9  & Entire file in \texttt{J} \\
blast.10 & 0-26~GB in \texttt{A}, 26-52~GB in \texttt{E}, remainder in \texttt{H} \\
blast.11 & 0-26~GB in \texttt{C}, 26-52~GB in \texttt{F}, remainder in \texttt{I} \\
blast.12 & 0-26~GB in \texttt{D}, 26-52~GB in \texttt{G}, remainder in \texttt{J} \\
blast.13 & 0-26~GB in \texttt{A}, 26-52~GB in \texttt{C}, remainder in \texttt{D} \\
blast.14 & 0-26~GB in \texttt{E}, 26-52~GB in \texttt{F}, remainder in \texttt{G} \\
blast.15 & 0-26~GB in \texttt{H}, 26-52~GB in \texttt{I}, remainder in \texttt{J} \\
blast.16 & 0-26~GB in \texttt{A}, 26-52~GB in \texttt{F}, remainder in \texttt{J} \\
blast.17 & 0-20~GB in \texttt{A}, 20-40~GB in \texttt{F}, remainder in \texttt{J} \\
blast.18 & 0-8~GB in \texttt{A}, 8-28~GB in \texttt{F}, remainder in \texttt{J} \\
\end{tabular}
\end{table}





From an I/O perspective, the first phase stores the query sequence in memory, memory maps the target database and sequentially scans the target database finding the seeds. The second phase operates completely in memory to combine the seeds where possible using in-memory data structures. The third phase can be viewed as ``random reads'' due to the non-uniform location of the seeds within the target database.

For these initial experiments, the \texttt{blastn} nucleotide-nucleotide algorithm from the NCBI C++ Toolkit version 2.2.29 was used. 
The target database is the \texttt{other\_genomic} nucleotide database distributed and pre-formatted by NCBI into 79 database fragments with each \texttt{.nsq} file no larger than 1~GB in size\cite{other_genomic}. There are $85,745$ sequences and $261,601,933,306$ total bases in the database. The queries are randomly sampled from the Reference CAMERA Viral Nucleotide Sequence database, \texttt{10570.V10}~\cite{10570}.



Table~\ref{tab:blast_benchmarks} lists the 18 striping patterns that were evaluated utilizing the Lustre directory types described earlier in Table~\ref{tab:directory_types}. Each benchmark was executed three times ``back-to-back-to-back'' using 64 processes spread across 8 nodes.

\subsection{Experimental Systems}

\subsubsection{knot}

The Lustre testbed, knot, consists of 35 Dell R720 servers, two Mellanox SX6036 FDR Infiniband switches, and a SAN. 
All nodes withn the testbed run the latest version of CentOS 6. 
Each server has two Intel Xeon 2630v2 processors running at 2.6Ghz.
All nodes have a Mellanox ConnectX-3 single port FDR Infiniband HCA and two 500GB 7200RPM SATA hard drives. 
Nineteen nodes are used as clients and have 128GB 1600MHz DDR3 RAM. 
Clients run Lustre 2.5.2 built with Mellanox OFED 2.2.
One of the client nodes is used as a login node, two are large-memory nodes (384GB RAM each), and 
the remaining sixteen client nodes are used as compute nodes for the tait compute cluster.
Six nodes are used as LNET routers and include an additional Connect3 FDR HCA. 
Two nodes are used as MDS/MGS nodes in a fail-over pair with six 15,000 RPM enterprise SAS drives used for MDT storage. 
The remaining eight nodes are used as OSS nodes, each having a quad-port 8Gb Qlogic fibre channel HBA connected to the SAN.

The SAN is composed of 10 pairs of LSI 2680 controllers. 
Each controller pair frontends two shelves of twelve drives each. 
Each OSS interfaces with one pair of controllers. 
The drives are presented as a single LUN, and ZFS is used to aggregate the individual drives that compose an OST. 
Multiple fibre channel paths are used for performance and redundancy.

In our 64 OST configuration, an OSS has 8 OSTs and each OST is composed from three drives. 
The six LNET routers have one connection to each SX6036. 
One SX6036 has connections to tait, and the other has connections to the Lustre server nodes.
The routers and servers are running Lustre 2.5.3 with ZFS 0.6.3. 

\subsubsection{Titan and Spider II}

Titan~\cite{titan} is an hybrid-architecture Cray XK7 system that contains 18,688 nodes connected with Cray's high-performance Gemini interconnect.
Each node features a 16-core AMD Opteron CPU, an NVIDIA K20 GPU with 6 GB DDR5 RAM, and 32 GB of main memory. 
In sum, Titan has 299,008 CPU cores, 18,688 GPUs, and a total system memory of 710 terabytes.

The Oak Ridge Leadership Computing Facility's center-wide Lustre deployment, called Spider II,
consists of four filesystems backed by four identical clusters. 
Each cluster consists of 72 OSS nodes serving storage provided by nine DDN SFA12K40 couplets.
Each couplet has two SFA12K40 controllers, ten SS7000 60-slot enclosures,
and 560 2 TB Near-Line SAS drives.
Additional details on the storage clusters is available in~\cite{sarp-cug}.

Titan uses 432 Cray XIO nodes as LNET routers, each of which can sustain 2880 MB/s of traffic.
The routers connect to one of 36 FDR Infiniband switches, and each switch contains an uplink to each of two 
108-port aggregation switches. 
The aggregation switches have direct connections to all of the MDS, MGS, OSS, and
file system management servers.

\section{Experimental Results}

\subsection{IOR}

\begin{figure}[t]
\centering
  \includegraphics[width=1.0\columnwidth]{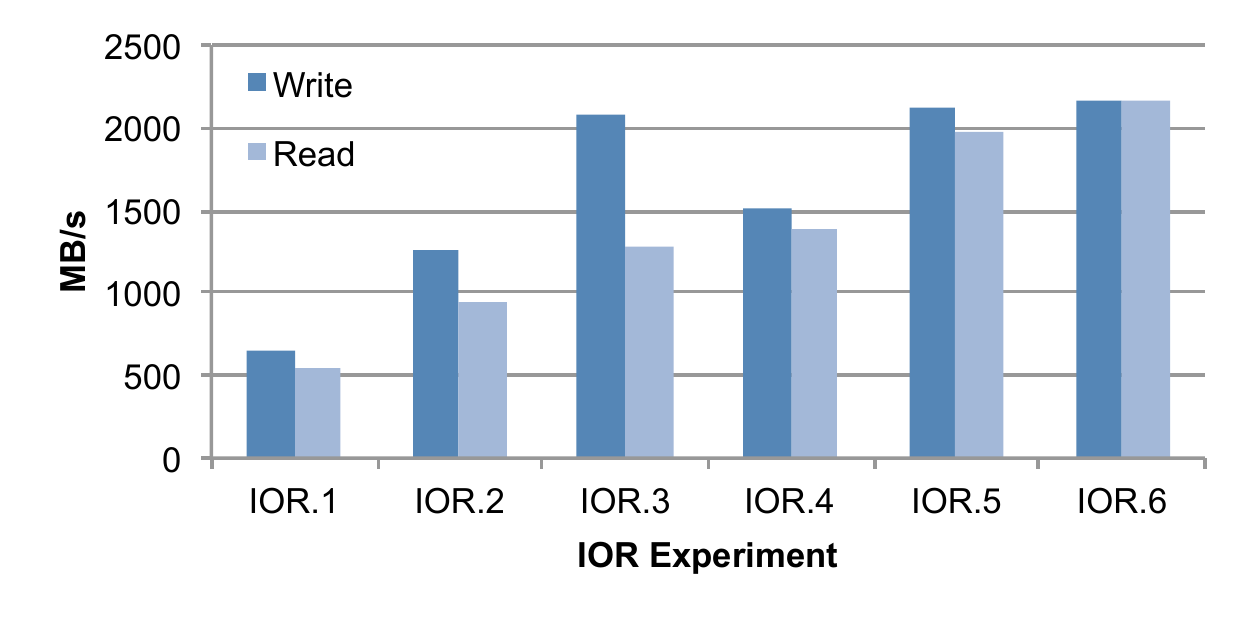}
\caption{IOR benchmark throughput results}
\label{fig:ior_results}
\end{figure}

Figure~\ref{fig:ior_results} presents the median write and read throughput for the various IOR benchmark experiments described earlier. Sequential read performance benefits significantly from dynamic striping (e.g., \emph{IOR.4-6}) compared to the static striping configurations (e.g., \emph{IOR.1-3}). Additionally, write performance with dynamic striping is equal to or greater when the minimum dynamic stripe count is at least equal to the static stripe count being compared (i.e., \emph{IOR.4} vs \emph{IOR.1}, \emph{IOR.5} vs \emph{IOR.2}, \emph{IOR.6} vs \emph{IOR.3}).

\begin{figure}[t]
\centering
  \includegraphics[width=1.0\columnwidth]{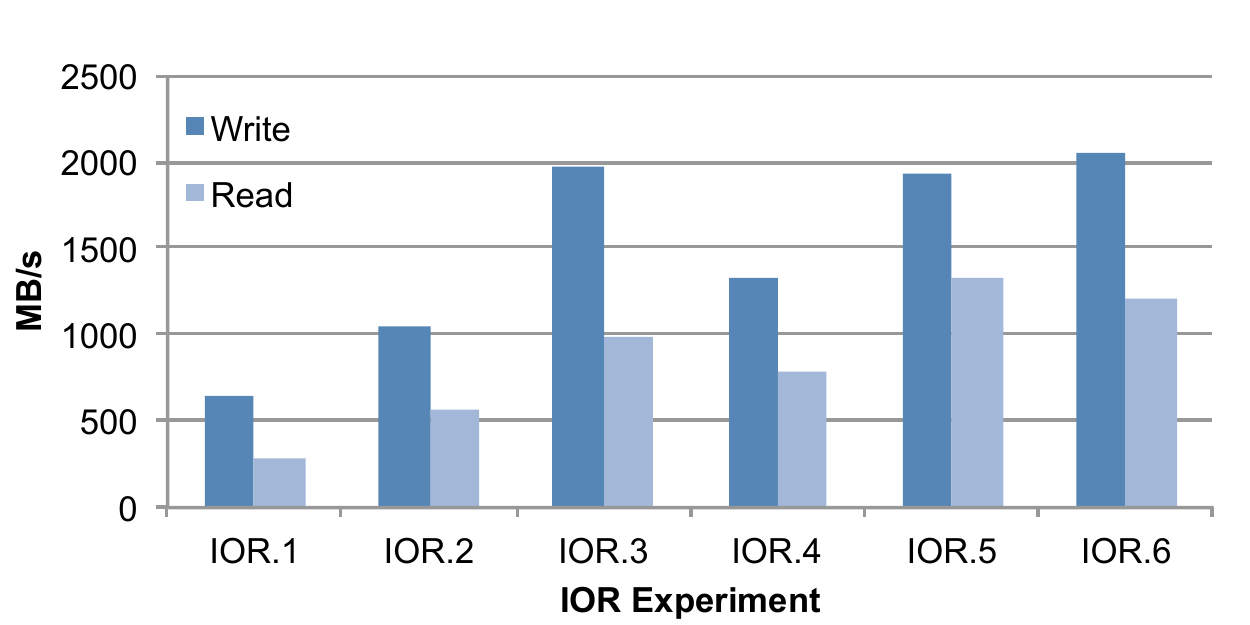}
\caption{IOR benchmark throughput results with ZFS cache disbled}
\label{fig:ior_no_zfs_results}
\end{figure}

Figure~\ref{fig:ior_no_zfs_results} presents the median write and read throughput for the same set of experiments described above, but with the ZFS caching disabled on the OST's. As expected, compared to the cache enabled results, the write performance is decreased slightly, around 7\% on average, while the read performance is significantly decreased by around 38\% on average. As with the IOR tests with ZFS caching enabled, both read and write performance for the simulated dynamic striping is equal to or better than the static striping where the initial dynamic stripe count stripe is equal to the static stripe count value. The anomaly in these results is the \emph{IOR.6} read performance being lower than the \emph{IOR.5} read performance, given that the tests with ZFS caching enabled indicated the opposite relationship. Since \emph{IOR.5} has the first terabyte of data stored with a stripe count of 4 and the remaining 3 terabytes stored with a stripe count of 16, while \emph{IOR.6} has the second terabyte stored at a stripe count of 8, the lower read throughput indicates that disabling the ZFS cache is resulting in a more true measure of the simulated dynamic striping performance. Disabling the ZFS cache was necessary, as our attempts to mitigate caching effects by simply reading unrelated large files between experiments failed to fool ZFS's intelligent adaptive replacement caching.

\subsection{netflow}

\begin{figure}[t]
\centering
  \includegraphics[width=1.0\columnwidth]{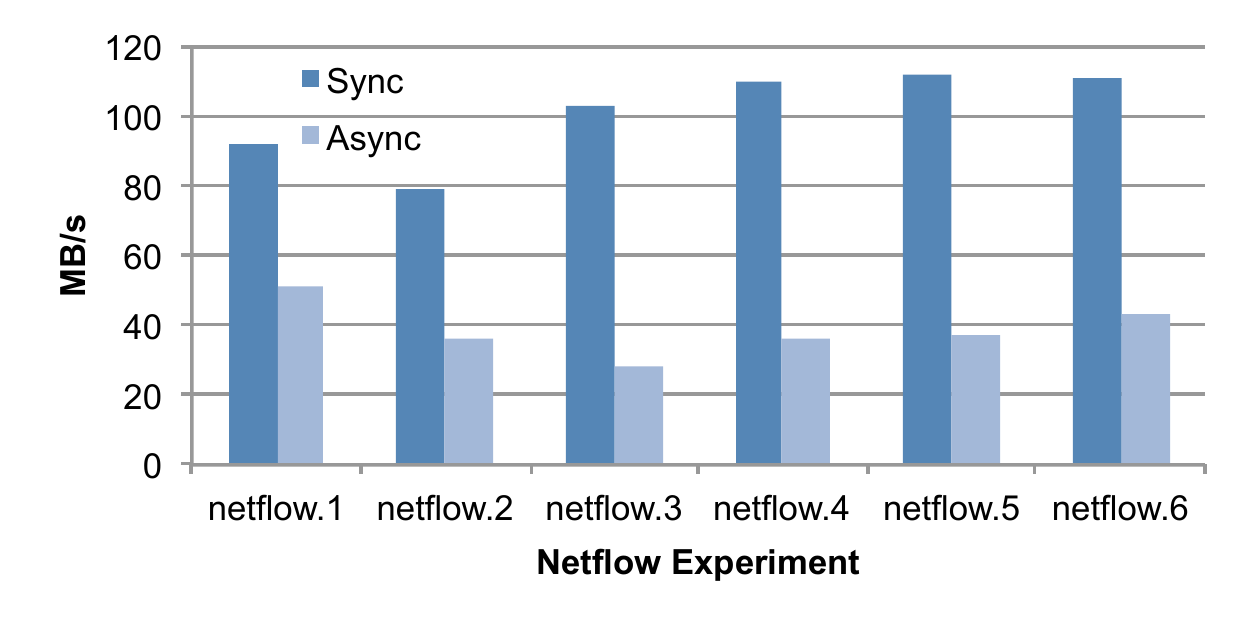}
\caption{Netflow phase 1 results - median read throughput}
\label{fig:netflow_phase1_results}
\end{figure}

Figure~\ref{fig:netflow_phase1_results} presents the median read performance from phase 1 of the netflow analysis workload. For the synchronous variation of the netflow workload, the performance for the simulated dynamic striping was better for all configurations than the static striping. Given that phase 1 of the netflow workload consists of large sequential reads, it is not surprising that these results are similar to the IOR read results. The asynchronous variation demonstrated much lower performance performance partially due to issues removing computation time when attempting to record the start time of the first I/O operation and the end time of the last I/O operation, and as such, should be given less weight for evaluation. 

\begin{figure}[t]
\centering
  \includegraphics[width=1.0\columnwidth]{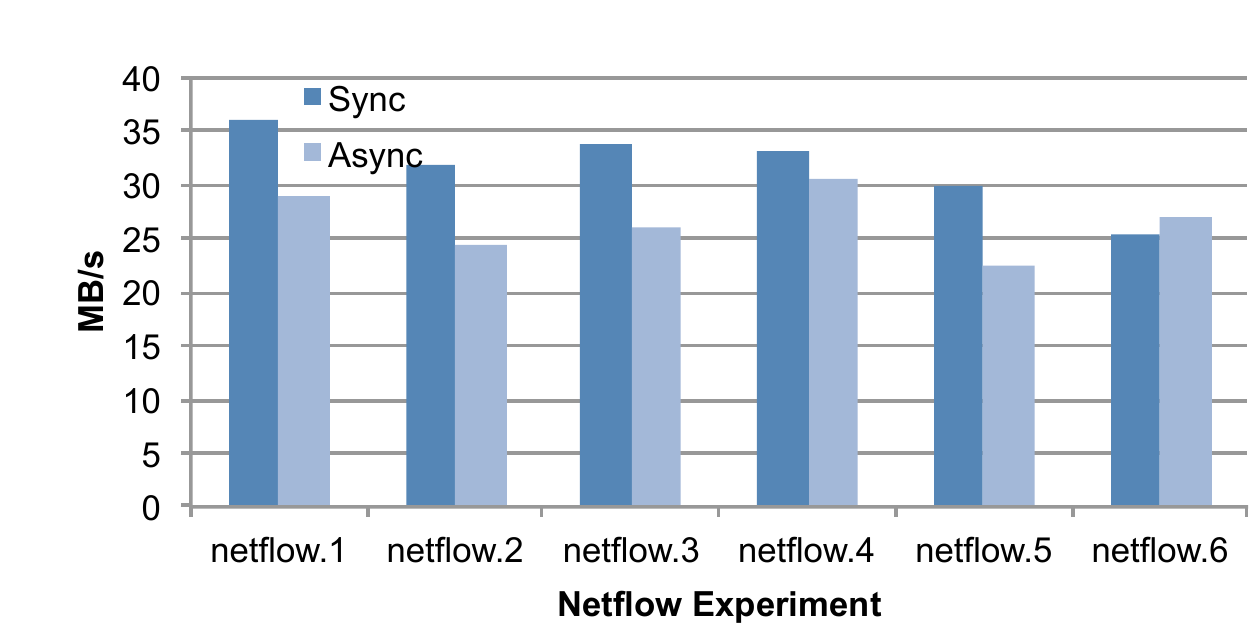}
\caption{Netflow phase 2 results - median read throughput}
\label{fig:netflow_phase2_results}
\end{figure}

Figure~\ref{fig:netflow_phase2_results} presents the median read performance from phase 2 of the netflow analysis workload.   With phase 2 consisting of some 70 million very small, random reads, the overall throughput was expected to be much less than seen in phase 1, but it was not expected that some of the simulated dynamic striping configurations would perform worse than their similar static counterparts.  These results indicate that analysis workloads characterized by large numbers of small, random reads may benefit from a smaller Lustre stripe count. 

\begin{figure}[t]
\centering
  \includegraphics[width=1.0\columnwidth]{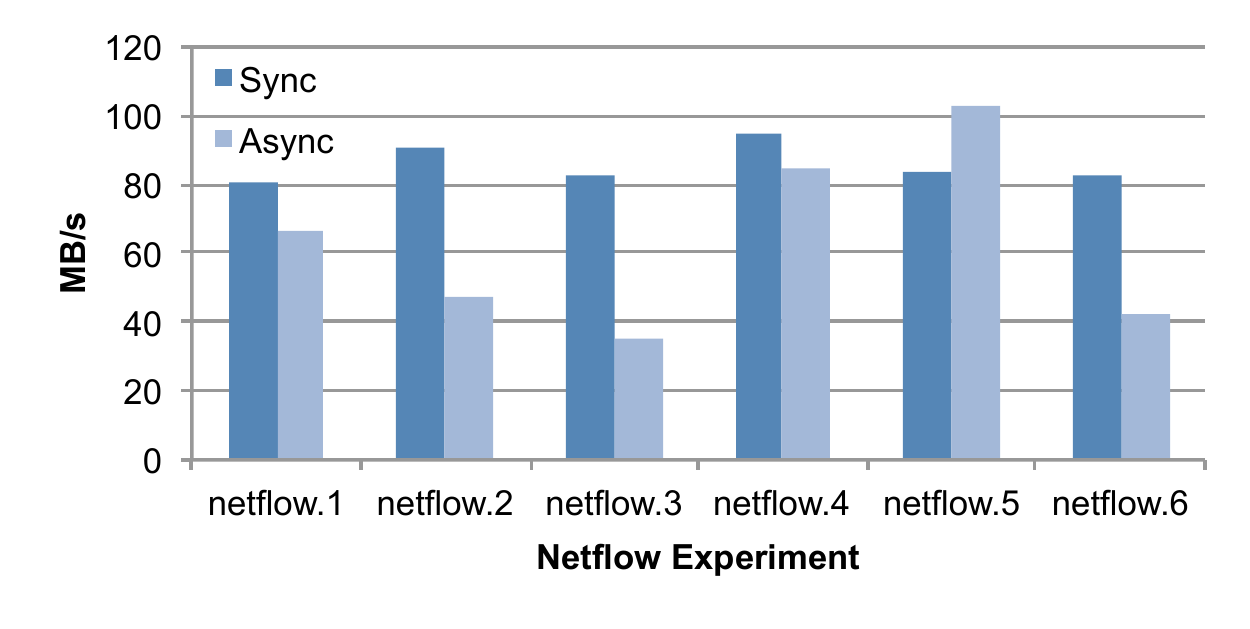}
\caption{Netflow phase 1 results with ZFS cache disabled - median read throughput}
\label{fig:netflow_phase1_no_zfs_results}
\end{figure}

Figure~\ref{fig:netflow_phase1_no_zfs_results} presents the results of the same netflow phase 1 test presented in figure~\ref{fig:netflow_phase1_results}, but with the ZFS caches disabled on the OST's. Disabling the ZFS caches decreased performance for the synchronous implementation by around 15\%, but the asynchronous performance indicated little difference. The synthetic dynamic striping improved performance in most cases compared to the static striping, but not as dramatically as the IOR read benchmark. This is most likely due to the much small file size (55 GB vs 4 TB for IOR), and the unaligned I/O caused by the netflow workload implementation when reading the variable length netflow records. 

\begin{figure}[t]
\centering
  \includegraphics[width=1.0\columnwidth]{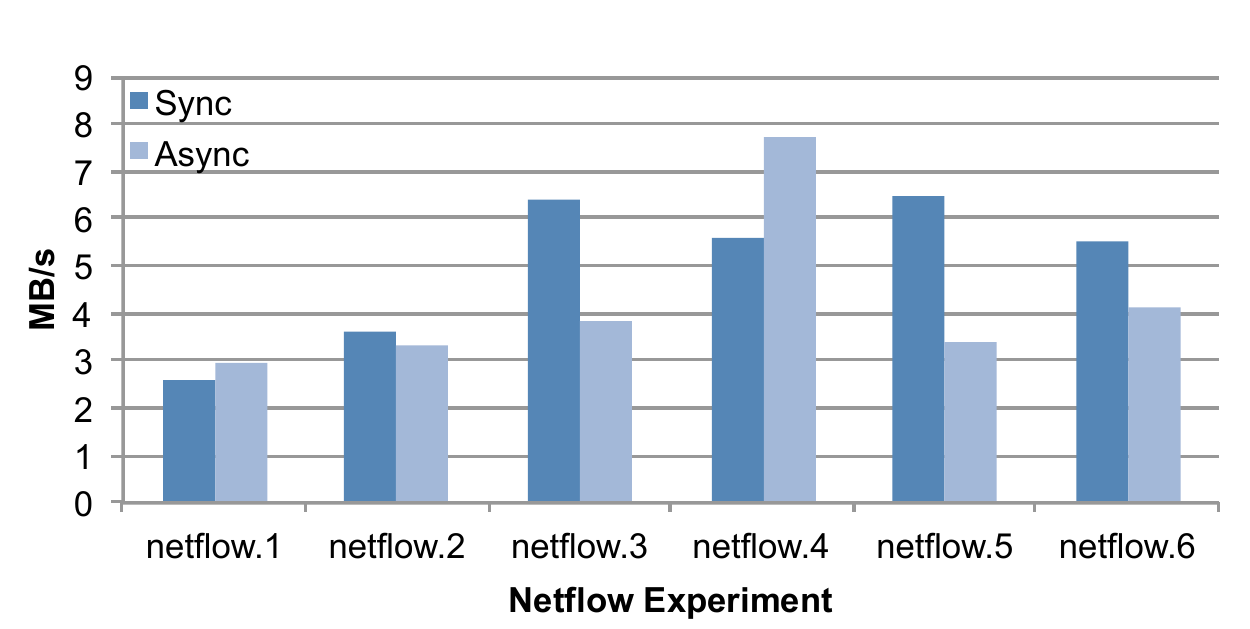}
\caption{Netflow phase 2 results with ZFS cache disabled - median read throughput}
\label{fig:netflow_phase2_no_zfs_results}
\end{figure}

Figure~\ref{fig:netflow_phase2_no_zfs_results} presents the results of the same netflow phase 2 test presented in figure~\ref{fig:netflow_phase2_results}, but with the ZFS caches disabled on the OST's. Disabling the ZFS caches decreased throughput for the synchronous implementations by around 26\% on average and by about 22\% on average for the asynchronous implementation. Clearly, the ZFS cache was of great benefit to the phase 2 random I/O when it was enabled in the earlier presented series of tests. The overall best performance was the \emph{netflow.4} configuration with the asynchronous variation of the netflow workload. A static striping configuration with a stripe count of 16 (\emph{neflow.3}) demonstrated the best performance for the synchronous variation of the netflow workload, but the \emph{netflow.5} dynamic configuration was very close. The throughput for all static and simulated dynamic configurations was similar, indicating that the performance of a large quantity of random reads is not significantly impacted by dynamic striping.

\subsection{BLAST}

\begin{figure}[t]
\centering
  \includegraphics[width=1.0\columnwidth]{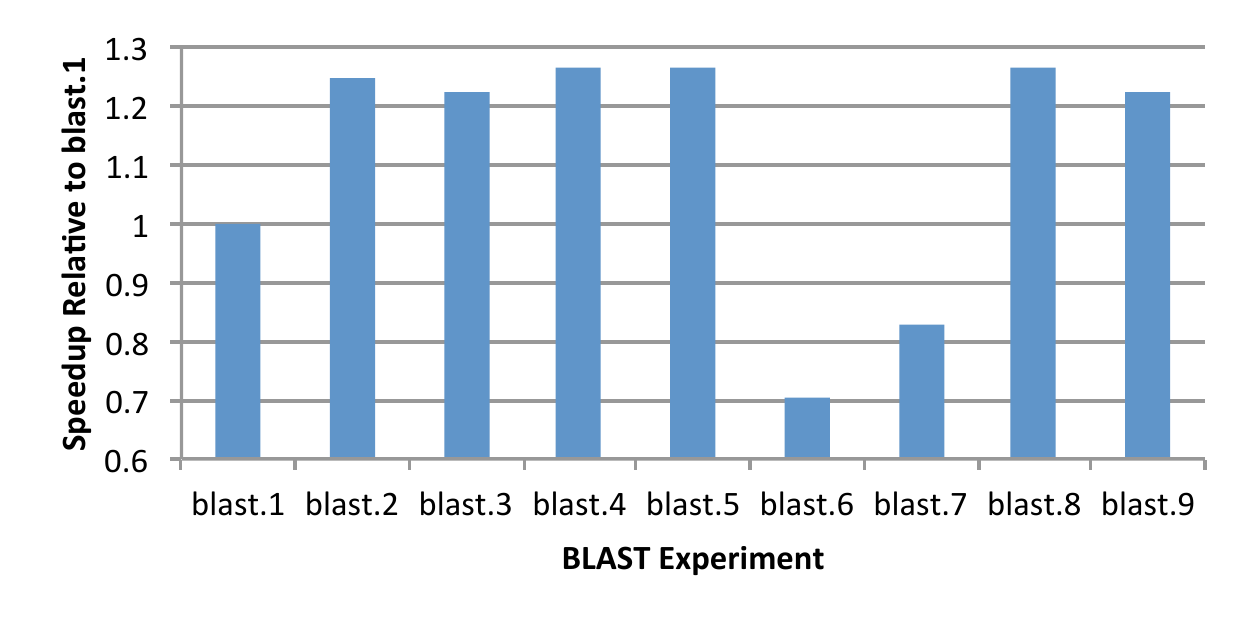}
\caption{BLAST results without dynamic striping - median speedup relative to default striping}
\label{fig:blast_1-9_results}
\end{figure}

\begin{figure}[t]
\centering
  \includegraphics[width=1.0\columnwidth]{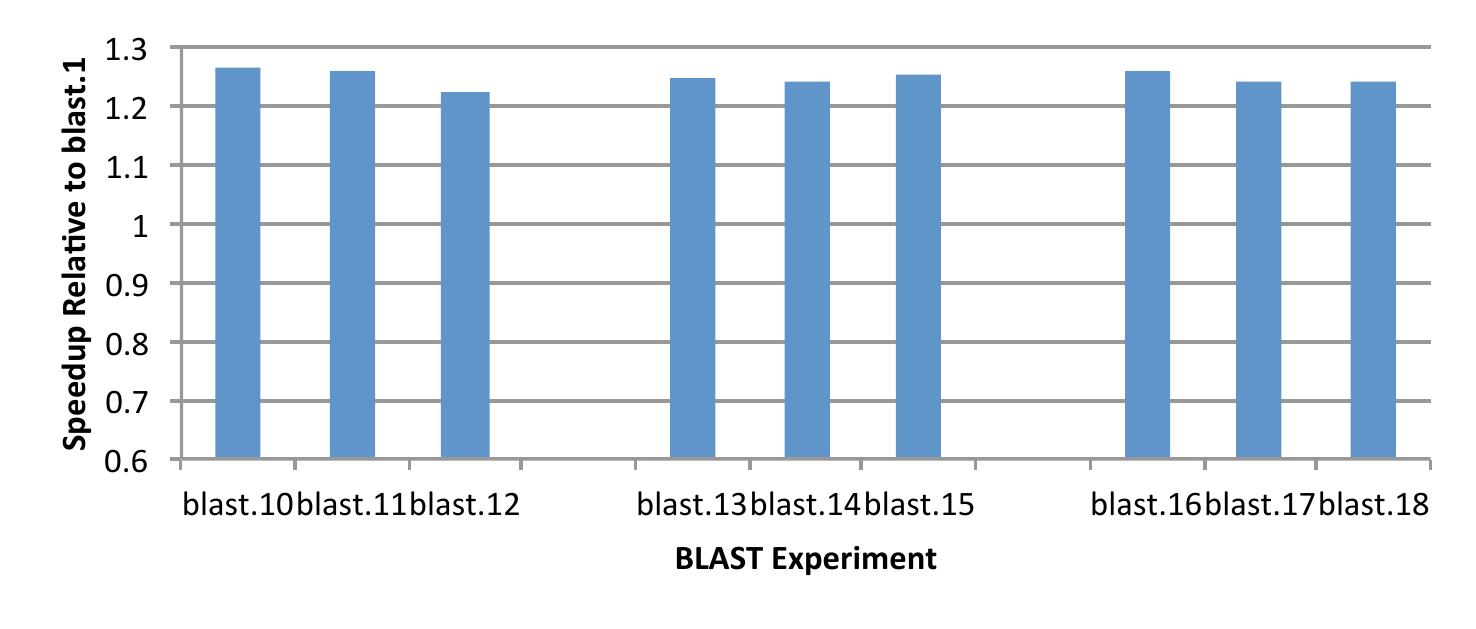}
\caption{BLAST results with dynamic striping - median speedup relative to default striping}
\label{fig:blast_10-18_results}
\end{figure}

Figure~\ref{fig:blast_1-9_results} and Figure~\ref{fig:blast_10-18_results} present the median speedup performance of the BLAST benchmarks, with and without dynamic striping, respectively, relative to the Titan/Spider default striping pattern (stripe count: 4 and stripe width: 1~MB) used by \emph{blast.1}. 

Excluding \emph{blast.6} (stripe count: 64 and stripe width: 2~MB) and \emph{blast.7} (stripe count: 4 and stripe width: 4~MB), the different static striping combinations (\emph{blast.2-9} in Figure~\ref{fig:blast_1-9_results}) provide a 1.22x-1.26x speedup. Since Titan is a shared resource within the OLCF and the \emph{blast.6} and \emph{blast.7} benchmarks were run in succession, the poor performance could be explained by a separate I$/$O-intensive job reading and writing to Spider at the same time.

When using dynamic striping, as shown in Figure~\ref{fig:blast_10-18_results}, fixing the stripe count and increasing the stripe width at each successive watermark (\emph{blast.10-12}) shows a slight performance decrease from a 1.26x to 1.22x speedup as the stripe count increases from 4 to 64 OSTs. The 4x increase in payload size likely benefits the sequential-read first phase of the BLAST algorithm, but hinders the performance of the final phase by excessively transferring unneeded data. This negative performance impact is expected to increase as the number of nodes increases.

Fixing the stripe width and increasing the stripe count at each successive watermark (\emph{blast.13-15}) and changing the watermark locations (\emph{blast.16-18}) do not appear to consistently increase or decrease performance. We suspect that using eight Titan client nodes is insufficient to demonstrate the benefits of wider watermarks.

\begin{figure}[t]
\centering
  \includegraphics[width=1.0\columnwidth]{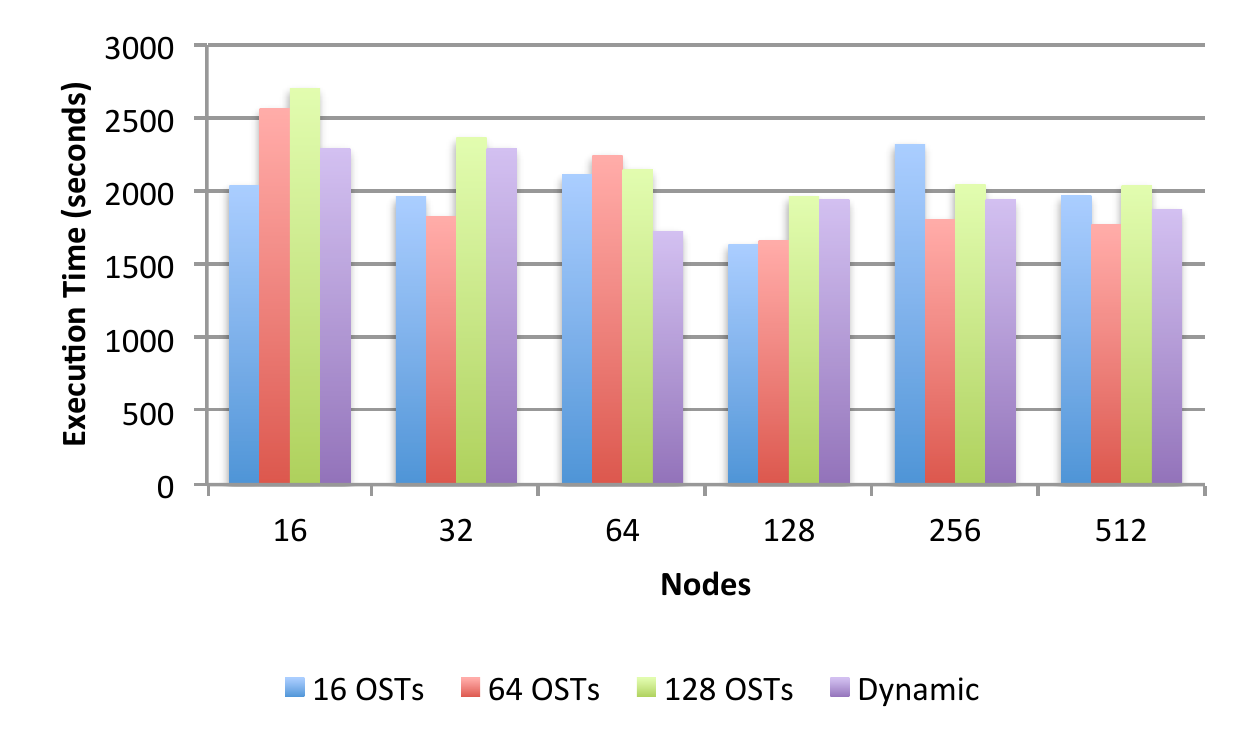}
\caption{Static vs Dynamic BLAST results using more nodes - median execution time}
\label{fig:blast_all_big_results}
\end{figure}

To evalute the impact of dynamic striping on larger numbers of nodes, various dynamic and static striping combinations were executed using 16, 32, 64, 128, 256, and 512 nodes on Titan. Figure~\ref{fig:blast_all_big_results} shows a representative subset of these tests with the stripe width set to 1~MB and the dynamic striping configuration set to have 0-26~GB in 16 OSTs, 26-52~GB in 64 OSTs, and the remainder in 128OSTs. Of particular interest is the fact that increasing the stripe count actually \emph{increases} execution time for each of these larger tests leading one to believe that larger stripe counts should be avoided. On the other hand, \emph{Dynamic} always outperforms \emph{128 OSTs} likely due to the ability to benefit from smaller stripe counts.

\section{Conclusions}

In this paper, we evaluated the ability to modify the Lustre striping characteristics of a file at different contiguous segments and call this new method, ``dynamic striping''. In our experiments, we focused on data analytic workloads using a watermark-based strategy where the stripe count and/or stripe width is modified once a file's size exceeds a one of the predetermined watermarks. The IOR benchmark results, along with the phase 1 netflow workload results, indicate that dynamic striping can benefit tasks with unpredictable data file size and large sequential reads. The BLAST workload and phase 2 of the netflow workload, with their significant random read phases, are more inconclusive, but did not demonstrate any compelling detrimental effects on performance. Rather, these results indicate that the optimal dynamic striping watermark and stripe width parameters are very dependent on the I/O pattern of the particular random read workload.  Overall, given the additional benefits of fault tolerance, adapting to OSS/OST faults by reducing stripe count, and easy expansion onto newly added OST's, these results demonstrate that dynamic striping would be a positive addition to the Lustre filesystem implementation.

\section{Future Work}

As part of our future work, we plan to run the netflow benchmarks with a much larger source data set. Separating phase 1 and phase 2 into to separate workloads to isolate the phase 2 results from any caching incurred during the phase 1 reads would result in more useful phase 2 results.

For the BLAST benchmarks, we plan to continue testing dynamic striping combinations on larger numbers of nodes on Titan while also testing even wider stripe counts. We also intend to execute the BLAST benchmarks on our Lustre testbed to obtain results in a contention-free environment. Collectively, these additional experiments should help reduce the inconsistency in the results arising from contention for a shared, center-wide file system.


\ifCLASSOPTIONcompsoc
  \section*{Acknowledgments}
\else
  \section*{Acknowledgment}
\fi

The authors would like to thank Brad Settlemyer for his efforts to conceptualize and direct the initial phase of this work. 
We thank Sarp Oral for his guidance and feedback on our experiments, Jeff Rossitter for providing a useful visualization tool based on parallel coordinates that helped explore multidimensional results, and Jason Hill for his expertise and efforts to help deploy our Lustre testbed.

This work used resources of the Extreme Scale Systems Center, supported 
by the Department of Defense, and the Oak Ridge Leadership Computing 
Facility, supported by the Office of Science of the Department of 
Energy, at Oak Ridge National Laboratory.



%

\end{document}